\documentclass[aps,prl,twocolumn,showpacs,preprintnumbers]{revtex4}

\usepackage{psfrag,graphicx}
\usepackage{dcolumn}
\usepackage{amsmath,amssymb}
\usepackage{bm}
\usepackage{amsfonts,amssymb,amsmath}        

\newcommand{\be}{\begin{equation}}
\newcommand{\ee}{\end{equation}}
\newcommand{\bq}{\begin{eqnarray}}
\newcommand{\eq}{\end{eqnarray}}

\newcommand{\ket}[1]{\left | \, #1 \right\rangle}

\begin{document}

\title{An Index Theorem for Graphene}
\date{\today}

\author{Jiannis K. \surname{Pachos$^1$} and Michael \surname{Stone$^2$}}
\affiliation{$^1$Department of Applied Mathematics and Theoretical Physics,
University of Cambridge, Wilberforce Road, Cambridge CB3 0WA, UK,\\
$^2$University of Illinois, Department of Physics 1110 W. Green St. Urbana,
IL 61801 USA.}

\begin{abstract}

We consider a graphene sheet folded in an arbitrary geometry, compact or
with nanotube-like open boundaries. In the continuous limit, the Hamiltonian
takes the form of the Dirac operator, which provides a good description of
the low energy spectrum of the lattice system. We derive an index theorem
that relates the zero energy modes of the graphene sheet with the topology
of the lattice. The result coincides with analytical and numerical studies
for the known cases of fullerene molecules and carbon nanotubes and it
extend to more complicated molecules. Potential applications to topological
quantum computation are discussed.

\end{abstract}


\maketitle

{\em Introduction:-} Much attention has been focused lately on various
geometric configurations of graphene, where an interplay takes place between
geometry and electronic properties such as its
conductivity~\cite{DiVincenzo,Tworzydlo,Gonzalez1,Gonzalez,Lammert,Kolesnikov,Novoselov}.
Previous methods for obtaining the zero modes of the system (electronic
eigenstates with zero energy) are based on lengthy analytical or numerical
procedures. As a possible alternative the much celebrated index
theorem~\cite{Atiyah} offers an analytic tool that relates the zero modes of
elliptic operators with the geometry of the manifold on which these
operators are defined. This theorem has a dramatic impact on theoretical and
applied sciences~\cite{Eguchi}. It allows to gain information about the
spectrum of widely used elliptic operators by simple geometric
considerations that could be otherwise hard or even impossible to determine.

It is the purpose of the present letter to establish a version of the index
theorem that relates the number of zero modes of graphene wrapped on
arbitrary compact surfaces to the topology of the surface. Nanotube-like
open boundaries that are of relevance to physical configurations of graphene
are also presented. When considering the low energy limit of graphene a
linearization of the energy is possible due to the presence of individual
Fermi points in the spectrum. This results in a Dirac equation defined on
the manifold of the lattice~\cite{Gonzalez1}, which describes the low energy
behavior of the system well. An additional coupling to an effective gauge
field with long range effect is generated by the deformations of the lattice
needed to introduce curvature. Since the Dirac operator is an elliptic
operator, it is possible to employ the index
theorem~\cite{Atiyah,Eguchi,Stone} to obtain information about the low
energy behavior of graphene and in particular about its conductivity
properties. Indeed, as we shall see in the following an exact relation can
be found that connects the number of zero modes with the genus, $g$, and the
number $N$ of possible open ends of the surface. Our results are in
agreement with the known cases of icosahedral fullerene
molecules~\cite{Kroto} and graphite nanotubes~\cite{Reich} where the
spectrum has been determined analytically or numerically. A relation between
the zero modes of more complicated molecules is provided.

There is a variety of applications that spring from this work. Information
about the spectrum of complex molecules constructed out of nanotubes can be
provided that may be impossible to obtain with other analytical approaches.
Moreover, the presence of $G$ fermionic zero modes dictates the existence of
a $2^G$ ground state degeneracy of the initial Hamiltonian. Hence, one could
employ reverse engineering and construct a fermionic lattice Hamiltonian
with a particular degeneracy structure. This is of much interest in the area
of topological quantum computation~\cite{Kitaev,Freedman}, where information
can be encoded in the degenerate states protected by topological
considerations. Moreover, lattice Hamiltonians such as the one that models
graphene can be engineered by Fermi atom gases superposed with optical
lattices~\cite{Experiments}. The detection procedure of the zero modes of
these models is well established~\cite{Ott} offering an alternative method
for probing the conductivity properties of various graphene configurations.
Similar approaches for the ground state degeneracy of fractional quantum
Hall systems in the planar case or on high-genus Riemannian surfaces have
been taken
in~\cite{Semenoff:1984dq,Schakel:1990mv,Jackiw1,Wen,Alimohammadi}.

\begin{center}
\begin{figure}[ht]
\resizebox{!}{3 cm}
{\includegraphics{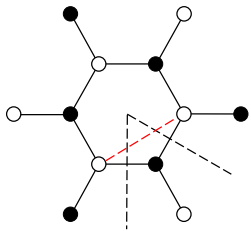} }
\caption{\label{lattice} The honeycomb lattice comprises of two
triangular lattices, A, denoted by black circles and, B, denoted by blank
circles. A single pentagonal deformation can be introduced by cutting a $\pi
/3$ sector and gluing the opposite sites together.}
\end{figure}
\end{center}

{\em The model:-} Let us first consider graphene, a flat isolated sheet of
graphite. It can be shown~\cite{Gonzalez} that the tight-binding
approximation reduces the system to that of coupled fermions on a honeycomb
lattice (see Fig.~\ref{lattice}). The relevant Hamiltonian is given by
\be
H = -t\sum_{<i,j>}a_i^\dagger a_j,
\label{Ham}
\ee
where $t>0$, $<\!\! i,j\!\!>$ denotes nearest neighbors on the lattice and
$a^\dagger_i$, $a_i$ are the fermionic creation and annihilation operators
at site $i$ with the non-zero anticommutation relation
$\{a_i,a_j^\dagger\}=\delta_{ij}$. The corresponding dispersion relation can
be easily evaluated as
\be
E(p) = \pm t\sqrt{1+ 4 \cos^2{\sqrt{3} p_y \over 2} + 4 \cos {3 p_x \over 2}
\cos {\sqrt{3} p_y \over 2} },
\label{dispersion}
\ee
where the interatomic distance is normalized to one. As can be deduced from
(\ref{dispersion}) at half-filling, graphene possesses two independent Fermi
points instead of Fermi lines. This rather unique property makes it possible
to linearize its energy by expanding it near the conical singularities of
the Fermi points. It is not hard to show that the resulting Hamiltonian is
given by the Dirac operator
\be
H_{\pm}= \pm{3 t\over 2} \sum_{\alpha=x,y}\gamma^\alpha p_\alpha,
\label{Dirac}
\ee
where the Dirac matrices, $\gamma^\alpha$, are given by the Pauli matrices,
$\gamma^\alpha=\sigma^\alpha$, and $\pm$ corresponds to the two independent
and oppositely positioned Fermi points. Hence, the low energy limit of
graphene is described by a free fermion theory. The corresponding spinors
are given by $(\ket{{\bf K}_\pm A},\ket{{\bf K}_\pm B})^T$ where A and B
denote the two different sublattices that comprise the honeycomb lattice
(see Fig.~\ref{lattice}) and ${\bf K_\pm}$ denote two independent Fermi
points chosen such that ${\bf K}_-=-{\bf K_+}$. One can see from
(\ref{Dirac}) that any fermionic mode rotated by $\sigma^z$ is mapped onto
another mode with the same energy and opposite momentum. This fact does not
necessarily hold for the zero modes of non-flat geometries as we shall see
in the following.

{\em Curvature deformations:-} In order to evaluate the effect of curvature
on the fermionic modes it is instructive to consider first the
transformation properties of the spinors. A $2\pi/3$ rotation of the lattice
centered on a hexagon is given by $R_{2\pi/3} = \exp (i {2\pi \over 3}
\sigma^z)$, which is an SU(2) element acting on the spinors, while a
$\pi$ rotation $R_{\pi} = \exp (i {\pi \over 2 } \sigma^z) i\tau^y$ mixes
both spinor elements and Fermi points, where $\tau^y$ is the Pauli operator
acting on the $K_\pm$ components~\cite{DiVincenzo,Lammert,Kolesnikov}. The
form of $R_\pi$ is due to a reorientation of the reference frame to its
original direction; it distributes an $i$ and a $-i$ to the spinor
components while the rotation of the Fermi momenta takes an additional minus
sign indicated by $i\tau^y$.

To obtain surfaces with arbitrary topology, curvature is introduced to an
initially flat honeycomb lattice by inserting deformations. In doing so, we
shall demand that each lattice site has exactly three neighbors and that the
lattice is inextensional (free to bend, impossible to stretch). The minimal
alteration of the honeycomb lattice that can introduce curvature without
destroying the cardinality of the sites is the insertion of a pentagon or a
heptagon; this corresponds to locally inserting positive or negative
curvature, respectively. Other geometries are also possible, leading to
similar results.

To introduce a single pentagon in a honeycomb lattice, one can cut a $\pi/3$
sector and glue the opposite sides together, as illustrated in
Fig.~\ref{lattice}. This causes no other defects in the lattice structure.
We shall demand that the spinors are smooth along the cut remedied by
introducing compensating fields which negate the
discontinuity~\cite{Lammert,Kolesnikov}. Indeed, the cut introduced in
Fig.~\ref{lattice} causes an exchange between A and B sublattices. This can
be incorporated into the Hamiltonian by introducing the non-abelian gauge
field, $\mathbf{A}$, with circulation
\be
\oint A_\mu dx^\mu = {\pi \over 2} \tau^y
\nonumber
\ee
that mixes the $+$ and $-$ spinor components. This flux can be attributed to
a fictitious magnetic monopole inside the surface with a charge contribution
of $1/8$ for each pentagon~\cite{Coleman}. In addition, moving a frame
around the deformation gives a non-trivial transformation that is equivalent
to a spin connection ${\bf Q}$. The flux of this field around the pentagon
is given by
\be
\oint Q_\mu dx^\mu = -{\pi \over 6}\sigma^z
\nonumber
\ee
and measures the angular deficit of $\pi/3$ around the cone. These fields
exactly compensate the spinor transformation $R_{\pi/3}=R_\pi
R_{2\pi/3}^{-1} =\exp (-i{\pi\over 6}\sigma^z) i\tau^y$ produced when the
lattice is rotated by $\pi/3$.

The modified Dirac equation, which incorporates the curvature and the
effective gauge field, couples the ${\bf K}_\pm$ spinor components together
due to the non-abelian character of ${\bf A}$. They can be decoupled by a
single rotation that gives
\be
{3t \over 2} \sum_{\alpha, \mu}\sigma^\alpha e^\mu_\alpha(p _\mu -iQ_\mu -i
A_\mu^k)\psi^k =E\psi^k,
\label{Dirac2}
\ee
where $k=1,2$ denotes the components in the rotated basis with $\oint
A^k_\mu dx^\mu =\pm\pi/2$. $e^\mu_\alpha$ is the zweibein of the curved
surface with metric $g_{\mu \nu}$ that defines the local flat reference
frame, $\eta_{\alpha \beta}=e^\mu_\alpha e^\nu _\beta g_{\mu\nu}$. This
equation faithfully describes the low energy behavior of graphene, such as
its zero modes, when it is deformed to an arbitrary surface.

{\em Index theorem:-} We have seen how our system reduces to the Dirac
equation of a spinor field on the surface of a lattice interacting with a
gauge field. Our aim now is to construct an index theorem that gives a
relation between the zero modes and the particular topology of the surface
on which the graphene sheet is wrapped.

Let us briefly review the index theorem. Consider a compact, oriented,
smooth, Riemannian manifold $M$ and the elliptic operator $D$ over $M$. Here
$D$ is the Dirac operator given in (\ref{Dirac2}). One can show that for
compact $M$ the Dirac operator is self-adjoint with a discrete spectrum of
eigenvalues~\cite{Esposito}. For even dimensional $M$, such as a surface,
the spinor space breaks up into two irreducible pieces, denoted by $V^+$ and
$V^-$. The Dirac operator interchanges between the two spaces in the
following way
\begin{eqnarray}
D&:& V^+ \rightarrow V^-
\\ \nonumber
D^*&:& V^- \rightarrow V^+.
\nonumber
\end{eqnarray}
Denoting by $\nu_{\pm}$ the dimension of the zero eigenspace of $V^\pm$, one
can define the index of $D$ by
\begin{equation}
\text{index}(D)\equiv \nu_+-\nu_-.
\nonumber
\end{equation}
One can introduce the operator $\gamma_5 \equiv i\prod_\alpha
\gamma^\alpha$ with eigenvalues $\pm 1$ that breaks the Hilbert space into
the two subspaces $V^+\oplus V^-$ with the property $\text{tr}
(\gamma_5)=\text{dim}(V^+)-\text{dim}(V^-) =\nu_+-\nu_-$. While there is a
spectral symmetry between the non-zero modes, the null-subspace does not
need to be symmetric due to topological defects, a fact that makes the
index$(D)$ a non-trivial quantity. Furthermore, the index
theorem~\cite{Atiyah,Eguchi,Stone,Esposito} states that
\begin{equation}
\text{index}(D)= {1 \over 2\pi} \int\!\!\! \int F,
\label{index_th}
\end{equation}
where $F=\partial\wedge A$ is the field strength of possible gauge
interactions and the integration is taken over the whole surface. Note that
there is no curvature contribution to the index theorem for two dimensional
manifolds.

Our aim is to evaluate the contribution from the gauge field, $F$, in
(\ref{index_th}). It is solely determined from the geometric characteristics
of the surface through the presence of pentagons and heptagons on the
original lattice. We shall only consider a surface that is either compact or
when open boundaries are present then a smooth differentiable surface can be
produced when the surface is ``glued" with its mirror symmetric one. One can
calculate the number of deformations in a lattice necessary to generate such
a surface by employing the Euler characteristic. Indeed, for V, E and F
being respectively the number of vertices, edges and faces of a lattice on a
surface with genus $g$, and $N$ open ends the Euler characteristic is given
by
\be
\chi = \text{V} - \text{E} + \text{F} = 2(1-g) -N.
\nonumber
\ee
We can easily verify that a single cut in the surface can reduce the genus
of the surface by one and increase the number of open ends by two , i.e.
$(g,N)\rightarrow(g-1,N+2)$, thus preserving the Euler characteristic,
$\chi$. From the total number of pentagons, hexagons and heptagons in the
lattice, $n_5$, $n_6$ and $n_7$, respectively, we see that E $= (5n_5 +6n_6
+ 7n_7)/2$, V$= (5n_5 +6n_6 + 7n_7)/3$ and F$=n_5 +n_6 + n_7$, giving
finally
\be
n_5-n_7=6\chi=12(1-g)-6N.
\label{Euler}
\ee
This reflects the fact that pentagons and heptagons have equal, but opposite
curvature and gauge flux contributions, while non-trivial topologies
necessarily introduce an imbalance in their numbers.

Eqn. (\ref{Euler}) reproduces the known case of a sphere ($\chi=2$, $n_5=12$
and $n_7=0$ for the C$_{60}$ fullerene), a torus ($\chi=0$, $n_5=n_7=0$ for
the nanotubes) or the genus-2 ($\chi=-2$, $n_5=0$, $n_7=12$) where equal
numbers of pentagons and heptagons can be inserted without changing the
topology of the surface.

Now we are in position to evaluate the index$(D)$. The contribution of the
gauge field term in~(\ref{index_th}) can be calculated straightaway from the
Euler characteristic. It is obtained by adding up the contributions from the
surplus of pentagons or heptagons. Thus, the total flux of the effective
gauge field can be evaluated by employing Stokes's theorem, giving
\be
{1 \over 2\pi} \int \!\!\!\!\int\! F ={1 \over 2\pi}\sum_{\text{def.}} \!
\oint A= {1\over 2\pi} (\pm{\pi\over 2}) (n_5-n_7)=\pm {3 \over 2}\chi,
\nonumber
\ee
where $\pm$ corresponds to the $k=1,2$ gauge fields. Hence, from
(\ref{index_th}), one obtains
\be
\nu_+-\nu_- = \left\{\begin{array}{cc}  \,\,\,\,\,3 \chi/2
,& \text{for $k=1$}\\
-3 \chi/2,& \text{for $k=2$} \end{array}\right..
\label{the_index}
\ee
As both of the cases contribute zero modes to the system the least number of
zero modes is given by $3|\chi| =|6-6g-3N|$, which coincides with their
exact number if $\nu_-=0$ or $\nu_+=0$.

This result reproduces the number of zero modes for the known molecules. The
fullerene, for which genus $g=0$ and $N=0$ has six zero modes which
correspond to the two triplets of C$_{60}$ and of similar larger
molecules~\cite{Gonzalez,Samuel}. For the case of nanotubes we have $g=0$
and $N=2$, which due to formula (\ref{the_index}) gives $\nu_+-\nu_-=0$.
This is in agreement with previous theoretical and experimental results
\cite{Saito,Reich}.

An alternative derivation of index$(D)$ for the case of nanotubes is given
when considering the operator $\gamma_5$, which in our case is given by
$\gamma_5=i\sigma^x\sigma^y=-\sigma^z$. As we have seen, for the case of
flat graphene sheets, transformations with respect to $\sigma^z$ give modes
with exactly the same energy even for the null subspace. When nanotubes are
considered then the dispersion relation (\ref{dispersion}) is accompanied by
the boundary condition $3 N_x p_x +\sqrt{3}N_y p_y = 2 \pi m$, where $N_x$,
$N_y$ determine the relative lattice position of the sites that are
identified, and $m$ is an integer~\cite{Saito}. One can easily see that the
$\gamma_5$ symmetry between the zero modes is still preserved giving, as
expected, zero difference between them. This symmetry also holds when
additional boundary conditions are employed to generate, e.g. the torus. In
the case of the fullerene, the topological defects cause the breakdown of
this symmetry.

{\em Conclusions:-} The presence or absence of zero modes in physical
systems is of much wider interest. For example, spin lattice models are
proposed~\cite{Kitaev1,Doucot} that exhibit ground state degeneracy that is
unaffected by small perturbations and, thus, capable of supporting error
free quantum information encoding. In our case $G$ fermionic zero modes
contribute $2^G$ ground state degeneracy, where $G$ depends on the topology
of the surface of the lattice. As we have seen, minimal local deformations
that do not change the topology can be introduced by having equal numbers of
pentagons and heptagons added in the lattice. These deformations do not
alter the number of zero modes given by the index theorem. Hence, this
methodology presents a promising way to construct topological models with
protected ground state degeneracy.

Finally, we would like to present an alternative physical model that can
realize Hamiltonian (\ref{Ham}). It consists of a single species ultra-cold
Fermi gas superposed with optical lattices in a honeycomb lattice obtained
by three planar standing wave lasers~\cite{Duan}. Recent
experiments~\cite{Experiments} demonstrate that it is possible to control
this system to a high degree of accuracy obtaining very low temperatures of
the order of $0.1T_F$ for arbitrary filling factors, where $T_F$ is the
Fermi temperature. At half filling one can realize the dynamics of
(\ref{Ham}) near the Fermi points, thus, simulating the conducting
properties of graphene in planar configuration. At this point one could
systematically study, for example, the effect of disorder. The latter can be
introduced either by deforming the geometry of the lattice, e.g. by
introducing pentagonal configurations at the boarders of the lattice, or by
considering impurities and lattice defects~\cite{Guinea}.

Detection of zero modes in similar systems has already been achieved in the
laboratory in the following way~\cite{Ott}. Consider the setup where the
Fermi gas is trapped with the optical lattice and an additional wide
harmonic potential. When the system is brought out of its equilibrium, i.e.
translated from the minimum, $x_{\text{min}}$, of the harmonic trap, then
two distinctive cases can arise depending on the presence or absence of zero
modes. The conducting regime is characterized by cloud oscillations around
$x_{\text{min}}$, while in the insulating case the center of mass of the
cloud remains at the displaced position exhibiting small Bloch oscillations.
This approach can offer an alternative experimental verification of the
relation between topological defects and the conductivity properties of
graphene.

{\em Acknowledgements:-} The authors would like to thank KITP for its
hospitality. This research was supported in part by the National Science
Foundation under Grant No. PHY99-0794 and by the Royal Society.

\end{document}